# Continuous variable polarization entanglement, experiment and analysis

Warwick P. Bowen,[1] Nicolas Treps,[1] Roman Schnabel,[1, 2] Timothy C. Ralph,[3] and Ping Koy Lam[1]

[1]*Department of Physics, Faculty of Science, Australian National University, ACT 0200, Australia*
[2]*Albert-Einstein-Institut, Universität Hannover, 30167 Hannover, Germany*
[3]*Department of Physics, Centre for Quantum Computer Technology,*
*University of Queensland, St Lucia, QLD, 4072 Australia*

We generate and characterise continuous variable polarization entanglement between two optical beams. We first produce quadrature entanglement, and by performing local operations we transform it into a polarization basis. We extend two entanglement criteria, the inseparability criteria proposed by Duan *et al.*[1] and the Einstein-Podolsky-Rosen paradox criteria proposed by Reid and Drummond[2], to Stokes operators; and use them to charactise the entanglement. Our results for the Einstein-Podolsky-Rosen paradox criteria are visualised in terms of uncertainty balls on the Poincaré sphere. We demonstrate theoretically that using two quadrature entangled pairs it is possible to entangle three orthogonal Stokes operators between a pair of beams, although with a bound $\sqrt{3}$ times more stringent than for the quadrature entanglement.

## INTRODUCTION

The ability to generate and manipulate pairs of photons that, when their polarisation is analyzed, demonstrate entanglement is a key tool of quantum optics. These states have allowed for many fundamental studies such as tests of Bells inequality[3]; and also perhaps more technologically minded studies like that of quantum computation[4]. It is surprising then that polarisation states in the other regime of quantum optics, that of continuous variables, have received comparatively little interest. Recently however, interest in continuous variable aspects of quantum polarisation states has been growing. This growth in interest is primarily due to the apparent applicability of continuous variable polarisation states to quantum information networks. It is generally accepted that a realistic quantum communications network must consist of nodes of atoms where quantum information algorithms are processed, linked by optical channels. In such a system optical quantum states must be transferable to the atomic nodes and visa versa. This quantum state transfer has been demonstrated between continuous variable polarisation states and spin states of an atomic ensemble[5]. Continuous variable polarisation states also do not require the network-wide local oscillator necessary when using other continuous variable states. This advantage, although technical, would result in a significant simplification of the infrastructure required for the network.

The polarisation state of light has four defining parameters, the Stokes parameters, one of which for a polarised beam, is redundant. This compares to two parameters, the quadrature amplitude and quadrature phase, for the most commonly studied continuous variable system. A number of theoretical papers on the generation and characterisation of polarisation squeezed states have now been published [6–9]; and several classes of polarisation squeezed states have been demonstrated experimentally[5, 10, 11]. Polarisation entanglement was introduced in the work of Korolkova *et al.*[9]. In their paper they suggest that polarisation entanglement may be generated by mixing a pair of polarisation squeezed beams on a 50/50 beam splitter, just as quadrature entanglement can be generated with a pair of quadrature squeezed beams. They also consider characterisation of polarisation entanglement; an proposing extension to the inseparability criterion introduced by Duan *et al.*[1] that is valid when the Stokes operators of interest are aligned orthogonally to the Stokes vector (see fig. 1), and an extension to the EPR paradox criterion introduced by Reid and Drummond[2].

This paper elaborates on our recent observation of polarisation entanglement[13]. It includes new experimental results on the Einstein-Podolsky-Rosen paradox, and more detailed discussion of most aspects of [13]. It should be noted that some of the theoretical background and preliminary experimental results presented here have been reported elsewhere[11, 12], we include them for completeness. We report the experimental transformation of entanglement between the phase and amplitude quadratures of two beams (quadrature entanglement)[14] onto a polarisation basis. This transformation is achieved by combining each quadrature entangled beam with an orthogonally polarised bright coherent beam on a polarizing beam splitter. Many method have been proposed to characterise quadrature entanglement, two commonly used criteria are the inseparability criterion proposed by Duan *et al.*[1, 15] and the EPR paradox criterion proposed by Reid and Drummond[2]. Both criteria are based explicitly on the uncertainty relation between the observables under interrogation. Using the standard uncertainty principle we generalise both criteria to an arbitrary pair of operators and then to Stokes operators in particular. The resulting EPR paradox criterion is identical to that given in [9]; our inseparability criterion however is, in contrast to the expression given in [9], valid for arbitrary Stokes vector orientation. We show that the polarisation entanglement we generate strongly satisfies both criteria. Interacting this entanglement with a pair of distant atomic ensembles could entangle the atomic spin states.

An interesting analogy may be made between continuous variable and discrete polarisation entanglement. Discrete polarisation entanglement is commonly observed between all three Stokes operators, and is basis independent. That is, correlations will exist between measurements on the two photons, when any arbitrary Stokes operator is measured. Since the entanglement discussed here is generated from a single quadrature entangled pair, in which entanglement is observed between only two quadratures, it is perhaps unsurprising that all three Stokes operators are not entangled. However, when two quadrature entangled pairs are utilised, we show that it is possible to simultaneously entangle all three Stokes operators, but only if the quadrature entanglement is strong enough to beat a bound $\sqrt{3}$ times stronger than that for the inseparability criterion. In contrast to discrete polarisation entanglement the entanglement is not basis independent. That is, observation of entanglement between three specific Stokes operators does not ensure entanglement between any three arbitrary Stokes operators.

## THEORY

### Polarisation and Stokes operators

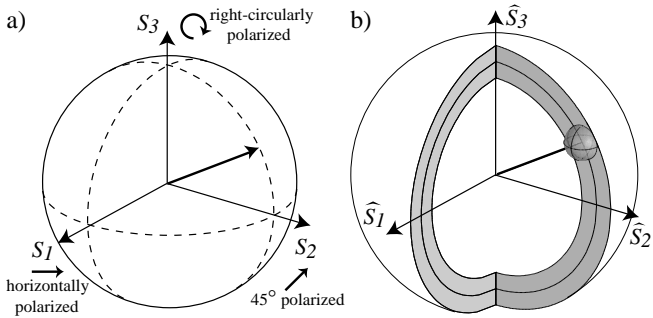

FIG. 1: Diagram of a) classical and b) quantum Stokes vectors mapped on a Poincaré sphere; the ball at the end of the quantum vector visualises the quantum noise in $\hat{S}_1$, $\hat{S}_2$, and $\hat{S}_3$; and the non-zero quantum sphere thickness visualises the quantum noise in $\hat{S}_0$.

In classical optics the polarisation state of a light beam can be described as a Stokes vector on a Poincaré sphere, as shown in fig. 1 a). It can be fully characterised by the four Stokes parameters[16]: $S_0$ measures the intensity of the beam; and $S_1$, $S_2$, and $S_3$ characterise its polarisation and form a cartesian axis system. If the Stokes vector points in the direction of $S_1$, $S_2$, or $S_3$, the polarised part of the beam is horizontally, linearly at $45°$, or right-circularly polarised, respectively. A pair of beams will not interfere if their Stokes vectors point in opposite directions. The radius of the classical Poincaré sphere is given by $S = (S_1^2 + S_2^2 + S_3^2)^{1/2}$ which describes the average intensity of the polarised part of the radiation. The degree of polarisation of a beam is given by the ratio of the intensity of the polarised part, to the total intensity $S/S_0$. For quasi-monochromatic laser light which is almost completely polarised $S_0$ is a redundant parameter, completely determined by the other three parameters ($S_0 = S$ in classical optics). The four Stokes parameters can be directly obtained from the mean value of the simple experiments shown in fig. 2.

The quantum mechanical Stokes operators are defined in much the same way as their classical counterparts. Following [26] we expand the Stokes operators in terms of the annihilation $\hat{a}$ and creation $\hat{a}^\dagger$ operators of the constituent horizontally (subscript H) and vertically (subscript V) polarised modes

$$\hat{S}_0 = \hat{a}_H^\dagger \hat{a}_H + \hat{a}_V^\dagger \hat{a}_V , \quad \hat{S}_2 = \hat{a}_H^\dagger \hat{a}_V e^{i\theta} + \hat{a}_V^\dagger \hat{a}_H e^{-i\theta}, \quad (1)$$
$$\hat{S}_1 = \hat{a}_H^\dagger \hat{a}_H - \hat{a}_V^\dagger \hat{a}_V , \quad \hat{S}_3 = i\hat{a}_V^\dagger \hat{a}_H e^{-i\theta} - i\hat{a}_H^\dagger \hat{a}_V e^{i\theta},$$

where $\theta$ is the phase difference between the H,V-polarisation modes. The commutation relations of the annihilation and creation operators

$$[\hat{a}_k, \hat{a}_l^\dagger] = \delta_{kl} , \quad \text{with} \quad k, l \in \{H, V\} , \quad (2)$$

directly result in Stokes operator commutation relations,

$$[\hat{S}_1, \hat{S}_2] = 2i\hat{S}_3 , \quad [\hat{S}_2, \hat{S}_3] = 2i\hat{S}_1 , \quad [\hat{S}_3, \hat{S}_1] = 2i\hat{S}_2 . \quad (3)$$

Apart from a normalisation factor, these relations are identical to the commutation relations of the Pauli spin matrices. In fact the three Stokes operators in Eq. (3) and the three Pauli spin matrices both generate the special unitary group of symmetry transformations SU(2)[17]. This group obeys the same algebra as the three-dimensional rotation group, so that distances in three dimensions are invariant. Therefore the operator $\hat{S}_0$ is also rotationally invariant and commutes with the other three Stokes operators ($[\hat{S}_0, \hat{S}_j] = 0$, with $j = 1, 2, 3$). The non-commutability of the other Stokes operators $\hat{S}_1$, $\hat{S}_2$ and $\hat{S}_3$ dictates the impossibility of the simultaneous exact measurement of their physical quantities; and even effects the definitions of the degree of polarisation[18, 19] and the Poincaré sphere radius. It can be shown from Eqs. (1) and (2) that the quantum Poincaré sphere radius is different from its classical analogue, $\langle \hat{S} \rangle = \langle \hat{S}_0^2 + 2\hat{S}_0 \rangle^{1/2}$. Furthermore, the noncommutability of the Stokes operators implies that entanglement of Stokes operators is possible between two beams, we term this continuous variable polarisation entanglement. Three observables are involved, compared to two for quadrature entanglement, and the entanglement between two of them relies on the mean value of the third.

The Stokes operators of a light beam can be characterised using the same apparatus as the classical Stokes parameters (fig. 2), and including an analysis of the fluctuations inherent in the measurement outcomes[9].

## Characterizing entanglement

Many techniques have been proposed to characterise continuous variable entanglement[20]. Since almost all continuous variable quantum optics experiments to date, including the ones reported here, have involved exclusively states with Gaussian noise statistics, we restrict ourselves to those states here. We utilise two common entanglement measures valid for Gaussian states; the inseperability criterion proposed by Duan *et al.*[1], and the Einstein-Podolsky-Rosen (EPR) paradox criterion proposed by Reid and Drummond[2]. In general, a necessary and sufficient criterion for entanglement should identify entanglement of any observables between a pair of sub-systems. Clearly however, a realistic criterion must be based on some finite set of observables. In this paper the term "polarisation entanglement" refers to entanglement that can be verified through measurements of only polarisation properties of the light field, and similarly "quadrature entanglement" refers to entanglement that can be verified through measurements performed solely on field quadratures.

Both the EPR and Inseparability criteria rely explicitly on the uncertainty relations between the observables involved and were initially proposed between the amplitude and phase quadratures of light beams. Given the uncertainty principle between an arbitrary pair of observables $\hat{A}$ and $\hat{B}$

$$\Delta^2 \hat{A} \Delta^2 \hat{B} \geq |\langle \delta \hat{A} \delta \hat{B} \rangle|^2 \qquad (4)$$
$$\geq \frac{|[\delta \hat{A}, \delta \hat{B}]|^2}{4} + \frac{|\langle \delta \hat{A} \delta \hat{B} + \delta \hat{B} \delta \hat{A} \rangle|^2}{4}$$

it is possible to generalise both criteria to any pair observables. Throughout this paper the variance of an operator $\hat{O}$ is calculated as $\Delta^2 \hat{O} = \langle \hat{O}^2 \rangle - \langle \hat{O} \rangle^2$, we expand all operators in terms of a mean unchanging term $\langle \hat{O} \rangle$ and a noise operator $\delta \hat{O}$ with zero expectation value, $\hat{O} = \langle \hat{O} \rangle + \delta \hat{O}$. Within some environment the commutation relation $[\delta \hat{A}, \delta \hat{B}]$ is a general property between the observables involved, and is independent of the properties of particular states; on the other hand the correlation function $|\langle \delta \hat{A} \delta \hat{B} + \delta \hat{B} \delta \hat{A} \rangle|^2$ measures correlation between the observables $\hat{A}$ and $\hat{B}$ for the particular state and is therefore dependent of the state properties. For this reason the correlation function term is generally neglected in the uncertainty relation. We neglect it here to obtain the standard uncertainty relation

$$\Delta^2 \hat{A} \Delta^2 \hat{B} \geq \frac{|[\delta \hat{A}, \delta \hat{B}]|^2}{4} \qquad (5)$$

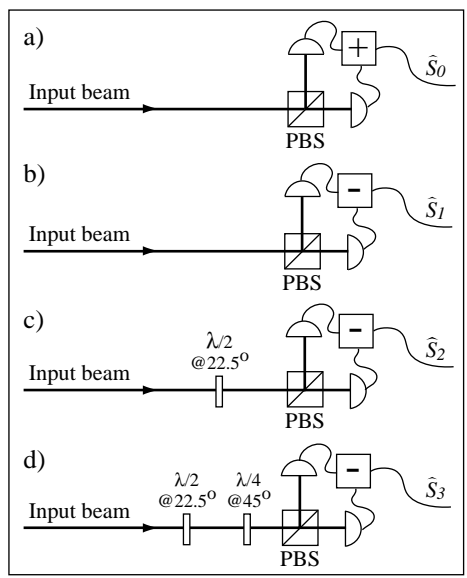

FIG. 2: Apparatus required to measure each of the Stokes parameters. PBS: polarizing beam splitter, $\lambda/2$ and $\lambda/4$: half- and quarter-wave plates respectively, the plus and minus signs imply that an electrical sum or difference has been taken.

When using this relation the entanglement criteria discussed herein become sufficient, but not necessary, for entanglement.

In section  we will consider the correlation function in particular for the states discussed explicitly and show that it has an insignificant contribution. Throughout this paper we label the two beams to be interrogated for entanglement with the subscripts $x$ and $y$ respectively. In general these beams will give different values for the correlation function. This leads to an ambiguous contribution to the uncertainty relation. We assume that beams $x$ and $y$ are interchangeable in the sense that all experimental outcomes are independent of their exchange. This is the situation relevant to our experiment and results in equal values of the correlation function for the two beams

$$|\langle \delta \hat{A} \delta \hat{B} + \delta \hat{B} \delta \hat{A} \rangle|^2 = |\langle \delta \hat{A}_x \delta \hat{B}_x + \delta \hat{B}_x \delta \hat{A}_x \rangle|^2$$
$$= |\langle \delta \hat{A}_y \delta \hat{B}_y + \delta \hat{B}_y \delta \hat{A}_y \rangle|^2 \qquad (6)$$

### The inseperability criterion

The inseparability criterion as originally proposed characterises the separability of the amplitude $\hat{X}^+$ and phase $\hat{X}^-$ quadratures of a pair of optical beams. For states with Gaussian noise statistics this criterion has been shown to be a necessary and sufficient criterion for entanglement[1].

The quadrature operators are observables and can be obtained from the annihilation and creation operators,

$$\hat{X}^+ = \hat{a} + \hat{a}^\dagger \qquad (7)$$
$$\hat{X}^- = i(\hat{a}^\dagger - \hat{a}) \qquad (8)$$

In the regime for which beams $x$ and $y$ are perfectly interchangeable and their fluctuations are symmetrical between the amplitude and phase quadratures the inseparability criterion can be written as

$$\Delta^2_{x \pm y} \hat{X}^+ + \Delta^2_{x \pm y} \hat{X}^- < 4 \qquad (9)$$

Throughout this paper $\Delta^2_{x \pm y} \hat{O}$ is the minimum of the variance of the sum or difference of the operator $\hat{O}$ between beams $x$ and $y$,

$$\Delta_{x\pm y}^2 \hat{O} = \min \langle (\delta \hat{O}_x \pm \delta \hat{O}_y)^2 \rangle \quad (10)$$

Given the Heisenberg uncertainty relation of equation 5 the measure in eq. (9) can be generalised to any pair of observables $\hat{A}$, $\hat{B}$

$$\Delta_{x\pm y}^2 \hat{A} + \Delta_{x\pm y}^2 \hat{B} < 2|[\delta \hat{A}, \delta \hat{B}]| \quad (11)$$

To allow direct analysis of our experimental results, we define the degree of inseparability $\mathcal{I}(\hat{A}, \hat{B})$, normalised such that if $\mathcal{I}(\hat{A}, \hat{B}) < 1$ the state is inseparable, and therefore entangled

$$\mathcal{I}(\hat{A}, \hat{B}) = \frac{\Delta_{x\pm y}^2 \hat{A} + \Delta_{x\pm y}^2 \hat{B}}{2|[\delta \hat{A}, \delta \hat{B}]|} \quad (12)$$

It should be noted that this measure may be generalised to a wider set of states by arranging it in a product form[21].

$$\mathcal{I}_{\mathrm{product}}(\hat{A}, \hat{B}) = \frac{\Delta_{x\pm y}^2 \hat{A} \Delta_{x\pm y}^2 \hat{B}}{|[\delta \hat{A}, \delta \hat{B}]|^2} \quad (13)$$

In this form the measure is independent of equal local squeezing operations performed on $x$ and $y$. Since the sum and product measures involve the same level of experimental complexity the product should, in general, be preferred. For our experimental configuration the measures are equivalent. Since the sum was the original form proposed by Duan *et al.*[1] we use it here.

### *The EPR paradox criterion*

The EPR paradox was first discussed by Einstein *et al.* as a demonstration of the physically unsatisfactory nature of quantum mechanics [22]. The EPR paradox criterion utilised here was proposed by Reid and Drummond[2] and is based on the observation of non-classical correlations between two beams. That is, the ability to infer (although not simultaneously) both variables of interest on beam $x$ to better than their Heisenberg uncertainty limit, after measurements on beam $y$. The EPR paradox criterion is a sufficient condition for entanglement and has been used to characterise entanglement in a number of experiments[14, 23–25]. It is given by

$$\Delta_{x|y}^2 \hat{X}^+ \Delta_{x|y}^2 \hat{X}^- < 1 \quad (14)$$

where $\Delta_{x|y}^2 \hat{O}$ is the variance of operator $\hat{O}$ in sub-system $x$ conditioned on its measurement in sub-system $y$ and is given by

$$\Delta_{x|y}^2 \hat{O} = \Delta^2 \hat{O}_x - |\langle \delta \hat{O}_x \delta \hat{O}_y \rangle|^2 / \Delta^2 \hat{O}_y \quad (15)$$

$$= \min_g \langle (\delta \hat{O}_x + g \delta \hat{O}_y)^2 \rangle \quad (16)$$

and the gain $g$ is an experimentally adjustable parameter. Utilizing the uncertainty relation of equation 5 this criterion can also be generalised to arbitrary observables

$$\Delta_{x|y}^2 \hat{A} \Delta_{x|y}^2 \hat{B} < \frac{|[\delta \hat{A}, \delta \hat{B}]|^2}{4} \quad (17)$$

Again, we express the criterion in terms of a factor, *the degree of EPR paradox* $\mathcal{E}(\hat{A}, \hat{B})$ normalised so that $\mathcal{E}(\hat{A}, \hat{B}) < 1$ implies observation of the EPR paradox.

$$\mathcal{E}(\hat{A}, \hat{B}) = 4 \frac{\Delta_{x|y}^2 \hat{A} \Delta_{x|y}^2 \hat{B}}{|[\delta \hat{A}, \delta \hat{B}]|^2} \quad (18)$$

**Generalisation of entanglement criteria to Stokes operators**

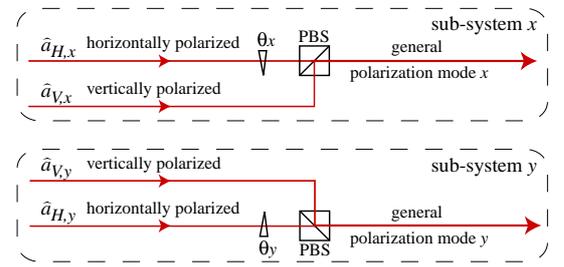

FIG. 3: Production of arbitrary polarisation modes. PBS: polarizing beam splitter.

From Eqs.(3) and (4) it is clear that the Stokes operator variances are restricted by the uncertainty relations[26]

$$\Delta^2 \hat{S}_1 \Delta^2 \hat{S}_2 \geq |\langle \hat{S}_3 \rangle|^2 + |\langle \delta \hat{S}_1 \delta \hat{S}_2 + \delta \hat{S}_2 \delta \hat{S}_1 \rangle|^2 / 4$$
$$\Delta^2 \hat{S}_2 \Delta^2 \hat{S}_3 \geq |\langle \hat{S}_1 \rangle|^2 + |\langle \delta \hat{S}_2 \delta \hat{S}_3 + \delta \hat{S}_3 \delta \hat{S}_2 \rangle|^2 / 4 \quad (19)$$
$$\Delta^2 \hat{S}_3 \Delta^2 \hat{S}_1 \geq |\langle \hat{S}_2 \rangle|^2 + |\langle \delta \hat{S}_3 \delta \hat{S}_1 + \delta \hat{S}_1 \delta \hat{S}_3 \rangle|^2 / 4$$

These uncertainty relations result in, typically, non-zero variances in the individual Stokes operators (see Fig. 1b)), and are ultimately what enables the verification of polarisation entanglement.

In general, any polarisation mode can be decomposed in terms of constituent horizontally and vertically polarised modes, with some phase angle $\theta$ between them. In this paper we consider a pair of arbitrary modes $x$ and $y$, decomposed in this way as shown in fig. 3. This inseparability criterion of eq. (13) was arrived at assuming that beams $x$ and $y$ were interchangeable; i.e. assuming that the outcome of any experiment in which they were involved would be independant of their exchange. In our experiment this condition is naturally satisfied, since our horizontally polarised modes are generated symmetrically on a 50/50 beam splitter, and the vertical constituents are identical coherent states. In order to satisfy the interchangeability condition assumed here, expectation values and variances of the horizontally (vertically) polarised input beams must be the same

$$\alpha_H = \langle \hat{a}_{H,x} \rangle = \langle \hat{a}_{H,y} \rangle \tag{20}$$

$$\alpha_V = \langle \hat{a}_{V,x} \rangle = \langle \hat{a}_{V,y} \rangle \tag{21}$$

$$\Delta^2 \hat{X}_H^\pm = \Delta^2 \hat{X}_{H,x}^\pm = \Delta^2 \hat{X}_{H,y}^\pm \tag{22}$$

$$\Delta^2 \hat{X}_V^\pm = \Delta^2 \hat{X}_{V,x}^\pm = \Delta^2 \hat{X}_{V,y}^\pm \tag{23}$$

and the relative phase between horizontally and vertically polarised modes for subsystems $x$ and $y$ must be related by $\theta = \theta_x = \pm\theta_y + m\pi$ where $m$ is an integer. Given these assumptions it is possible to calculate $\mathcal{I}(\hat{S}_i, \hat{S}_j)$ from eqs. (1) and (3). From consideration of our experimental setup, we find some further simplifications possible. We assume that the horizontal and vertical inputs are not correlated,

$$\langle \delta \hat{X}_{H,x/y}^\pm \delta \hat{X}_{H,x/y}^\pm \rangle = 0 \tag{24}$$

$$\langle \delta \hat{X}_{H,x/y}^\pm \delta \hat{X}_{H,x/y}^\mp \rangle = 0 \tag{25}$$

and that each input beam does not exhibit internal amplitude/phase quadrature correlations

$$\langle \delta \hat{X}_{H,x/y}^+ \delta \hat{X}_{H,x/y}^- + \delta \hat{X}_{H,x/y}^- \delta \hat{X}_{H,x/y}^+ \rangle = 0 \tag{26}$$

$$\langle \delta \hat{X}_{V,x/y}^+ \delta \hat{X}_{V,x/y}^- + \delta \hat{X}_{V,x/y}^- \delta \hat{X}_{V,x/y}^+ \rangle = 0 \tag{27}$$

Finally we assume that the vertically polarised input modes are bright ($\alpha_V^2 \gg 1$) so that second order terms are negligible. Given these assumptions the Stokes operator expectation values for both beams $x$ and $y$ are given by

$$\langle \hat{S}_1 \rangle = \alpha_H^2 - \alpha_V^2 \tag{28}$$

$$\langle \hat{S}_2 \rangle = 2\cos\theta \alpha_H \alpha_V \tag{29}$$

$$\langle \hat{S}_3 \rangle = 2\sin\theta \alpha_H \alpha_V \tag{30}$$

The degree of inseparability for each of the three permutations of Stokes operators is then given by

$$\mathcal{I}(\hat{S}_1, \hat{S}_2) = \frac{\Delta_{\pm y}^2 \hat{S}_1 + \Delta_{\pm y}^2 \hat{S}_2}{8|\sin\theta \alpha_H \alpha_V|} \tag{31}$$

$$\mathcal{I}(\hat{S}_1, \hat{S}_3) = \frac{\Delta_{\pm y}^2 \hat{S}_1 + \Delta_{\pm y}^2 \hat{S}_3}{8|\cos\theta \alpha_H \alpha_V|} \tag{32}$$

$$\mathcal{I}(\hat{S}_2, \hat{S}_3) = \frac{\Delta_{\pm y}^2 \hat{S}_2 + \Delta_{\pm y}^2 \hat{S}_3}{4|\alpha_H^2 - \alpha_V^2|} \tag{33}$$

and the degree of EPR paradox for each permutation is

$$\mathcal{E}(\hat{S}_1, \hat{S}_2) = \frac{\Delta_{x|y}^2 \hat{S}_1 \Delta_{x|y}^2 \hat{S}_2}{|\sin\theta \alpha_H \alpha_V|^2} \tag{34}$$

$$\mathcal{E}(\hat{S}_1, \hat{S}_3) = \frac{\Delta_{x|y}^2 \hat{S}_1 \Delta_{x|y}^2 \hat{S}_3}{|\cos\theta \alpha_H \alpha_V|^2} \tag{35}$$

$$\mathcal{E}(\hat{S}_2, \hat{S}_3) = 4 \frac{\Delta_{x|y}^2 \hat{S}_2 \Delta_{x|y}^2 \hat{S}_3}{|\alpha_H^2 - \alpha_V^2|^2} \tag{36}$$

Notice than when $\theta = m\pi$ where $m$ is an integer the denominators of eqs (31) and (34) become zero. It is then not possible to verify the presence of entanglement between $\hat{S}_1$ and $\hat{S}_2$. Since our measure is only a sufficient criterion for entanglement, this result does not exclude its existence, only our ability to measure it. In this situation a more detailed entanglement criterion, including the correlation function, is required. The same is true if $\alpha_H$ or $\alpha_V$ equal zero; between $\hat{S}_1$ and $\hat{S}_3$ when $\theta = (m + 1/2)\pi$; and between $\hat{S}_2$ and $\hat{S}_3$ when $|\alpha_H| = |\alpha_V|$. Section includes a discussion of this effect. In our experiment however, we took care to characterise the entanglement in regimes for which the measures of eqs. (12) and (18) are effective.

## EXPERIMENT

### Generation of quadrature squeezing

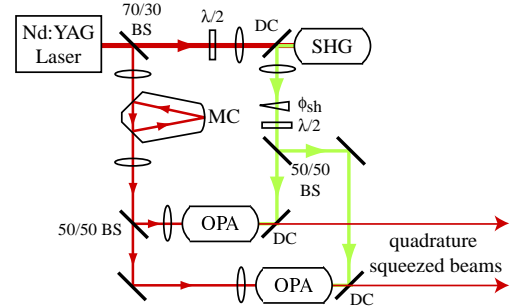

FIG. 4: Experimental apparatus used to generate two squeezed beams. BS: beam splitter, MC: mode cleaning resonator, DC: Dichroic, $\lambda/2$: half-wave plate, $\phi_{sh}$: second harmonic phase shifter.

In this work, quadrature entanglement was transformed into polarisation entanglement. We produced the quadrature entanglement from a pair of amplitude squeezed beams. The experimental setup that was used to generate these beams is shown in fig. 4. The laser source was a 1.5 W 1064 nm monolithic ring Nd:YAG laser. Roughly half of the output power was mode-matched into an intracavity second harmonic generator (SHG) consisting of a MgO:LiNbO$_3$ crystal with one flat dual anti-reflection coated surface and one 10 mm radius of curvature dual high-reflection coated surface, and a 25 mm radius of curvature output coupler with 92 % and 6% reflectivity for the fundamental and second harmonic light, respectively. 350 mW of second harmonic 532 nm light was produced. The SHG was locked with a Pound-Drever-Hall[27] technique based on an intra-cavity phase modulation introduced through a 29.7 MHz refractive index modulation on the MgO:LiNbO$_3$ crystal. As a consequence of this the second harmonic beam had a phase modulation at 29.7 MHz. The remainder of the 1064 nm light was mode-matched into a high-finesse ring resonator to remove laser relaxation oscillation noise at MHz frequencies. This resonator was locked using a phase sensitive spatial mode technique (Tilt locking) [28]. Part of the now spectrally cleaned beam was used to seed a pair of optical parametric amplifiers (OPAs) of similar construction to the SHG but with output coupler reflectivity of 96 % for 1064 nm light. The 532 nm light was used to pump both OPAs. Depending on the relative phase of the seed and pump beams each OPA output was an amplified or deamplified version of its seed. The phase modulation on the pump beams caused by the SHG locking modulated the amplification of the OPAs. This modulation was used to lock to either amplification or deamplification. When the OPAs were locked to deamplification the amplitude noise of the seed was also deamplified resulting in amplitude squeezed beams. When locked to amplification each OPA produced a phase squeezed beam. In this experiment we used amplitude squeezed beams, typical spectra for which are shown in fig. 5. These spectra were measured in a balanced homodyne detector with overall efficiency of approximately 85 %. The degradation of squeezing at low frequencies was due to the resonant relaxation oscillation of the laser, at high frequencies the squeezing was limited by the resonator bandwidth of our OPAs.

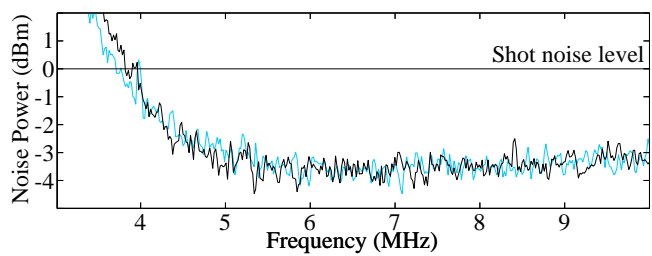

FIG. 5: Squeezing spectra observed from the two OPAs, normalised to the shot noise level.

**Generation and characterisation of quadrature entanglement**

Quadrature entanglement is commonly produced by interfering a pair of quadrature squeezed beams on a 50/50 beam splitter[14]. The relative phase of the two beams is of critical importance and, when using two amplitude squeezed beams, should be $\pi/2$. We interfered our amplitude squeezed beams with 97.8 % mode-matching efficiency. The relative phase between them was actively controlled to $\pi/2$ by balancing the carrier powers of the two output beams. To quantify the extent of the entanglement we mode-matched each entangled beam with 93 % efficiency into a balanced homodyne detector. Epitaxx ETX500 photodiodes also with 93 % efficiency were used. A phase modulation at 30.5 MHz on each entangled beam enabled, through a RF side-band locking technique, each homodyne detector to be locked to measure the amplitude quadrature of its entangled beam. The phase quadrature was measured by actively balancing the power splitting inside each homodyne detector. We observed correlations on both the amplitude and phase quadratures between the two entangled beams, which is a strong signature of quadrature entanglement. We quantified both the quadrature inseparability and EPR paradox criteria and obtained the results $\mathcal{I}(\hat{X}^+, \hat{X}^-) = 0.44 \pm 0.01$ and $\mathcal{E}(\hat{X}^+, \hat{X}^-) = 0.58 \pm 0.02$, which are both well below the limit of unity for entanglement. These results are discussed in detail in [25], including an experimental analysis of the effect of loss on each criteria, and an interpretation in terms of sideband photon numbers.

**Transformation to polarisation entanglement**

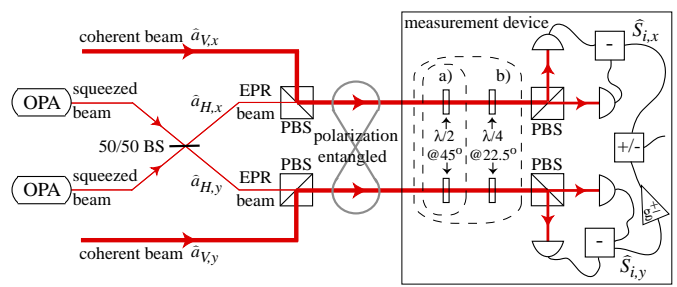

FIG. 6: Experimental production and characterisation of continuous variable polarisation entanglement. The optics within a) are included to measure $\hat{S}_2$, and those within b) to measure $\hat{S}_3$. BS: beam splitter, PBS: polarizing beam splitter.

We transformed the entanglement onto a polarisation basis by combining each quadrature entangled beam (horizontally polarised) on a polarizing beam splitter with a much more intense vertically polarised coherent beam ($\alpha_V^2 = 30\alpha_H^2$) (see fig. 6). The overlap efficiency between the modes was observed to be 91%, and the relative phase between the horizontal and vertical input modes $\theta$ was controlled to be $\pi/2$. In this situation the denominators of eqs. (32) and (35) are equal to zero, so that it is not possible to verify entanglement between $\hat{S}_1$ and $\hat{S}_3$. We therefore only characterise the entanglement criteria for the other two combinations of Stokes operators ($\hat{S}_1$ and $\hat{S}_2$; and $\hat{S}_2$ and $\hat{S}_3$).

*Individual characteristics of the two polarisation entangled beams*

The Stokes operators of each of the polarisation entangled beams were measured as shown in fig. 2. Initially each beam was split on a polarizing beam splitter and the two outputs detected with a pair of epitaxx ETX500 photodiodes. The sum of the two photocurrents gave a measurement of $\hat{S}_0$, and the difference, $\hat{S}_1$. With the inclusion of a half-wave plate before the polarizing beam splitter, the difference yielded an instantaneous value for $\hat{S}_2$, and with a quarter-wave plate $\hat{S}_3$ (see fig. 6). Fig. 7 presents variance spectra for all four Stokes operators of the polarisation entangled beams measured independently. All of the results presented in this paper were taken over the sideband frequency range from 2 to 10 MHz and are the average of ten consecutive traces. Every trace was more than 4.5 dB above the measurement dark noise which was taken into account. All four spectra exhibit high levels of noise at low frequencies. This is primarily due to resonant relaxation oscillation noise from our laser. The spectra in fig. 7 (a) display $\hat{S}_0$ for the two polarisation entangled beams. Each spectra is equivalent to the total intensity noise of the constituent coherent and quadrature entangled beams. Since the coherent beam was much brighter than the entangled beam its contribution to the spectra is dominant. This caused the spectra of $\hat{S}_1$ (fig. 7 (a)), being the difference of the intensity noise of the constituent beams, to be almost identical to those for $\hat{S}_0$. Both $\hat{S}_0$ and $\hat{S}_1$ display laser relaxation oscillation noise at low frequencies but become shot noise limited at frequencies above 5 MHz. The fact that they are shot noise limited implies that entanglement involving either is unlikely. Both the spectra for $\hat{S}_2$ and $\hat{S}_3$ are well above the shot noise throughout the measurement range. This is because the highly noisy fluctuations of the quadrature entangled beams has been mapped onto these operators, and suggests that they may be entangled.

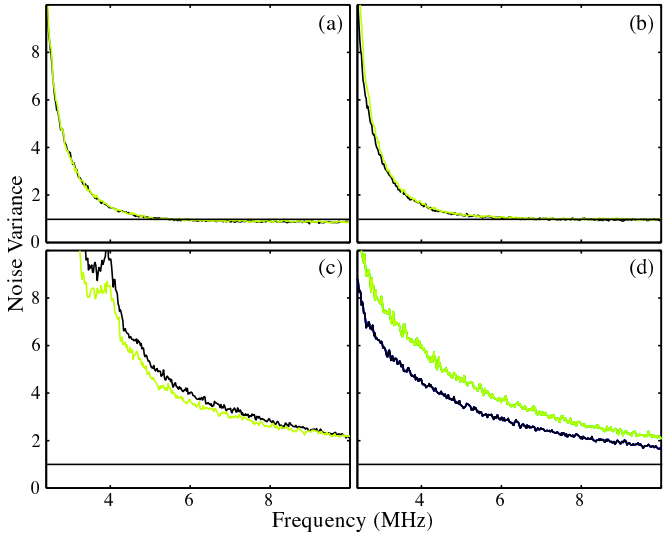

FIG. 7: Experimental measurement of the Stokes operator variances of the entangled beams, (a) $\Delta^2 \hat{S}_0$, (b) $\Delta^2 \hat{S}_1$, (c) $\Delta^2 \hat{S}_2$, (d) $\Delta^2 \hat{S}_3$; the black traces are for beam $x$ and the grey traces for beam $y$.

*Measurement of the inseparability criterion*

A clearer signature of entanglement was evident in the form of strong correlations of both $\hat{S}_2$ and $\hat{S}_3$ between the two resulting beams. We quantified the inseparability and EPR paradox criteria for entanglement of these beams. This quantifi-cation of $\mathcal{I}(\hat{S}_i, \hat{S}_j)$ and $\mathcal{E}(\hat{S}_i, \hat{S}_j)$ required measurements of $\alpha_V$, $\alpha_H$, $\Delta^2_{\pm}\hat{S}_i$, and $\Delta^2_{x|y}\hat{S}_i$. We determined $\alpha_V^2$ directly by blocking the horizontal modes and measuring the power spectrum of the subtraction between the two homodynes, this also gave $\alpha_H^2$ since the ratio $\alpha_V^2/\alpha_H^2$ was measured to equal thirty.

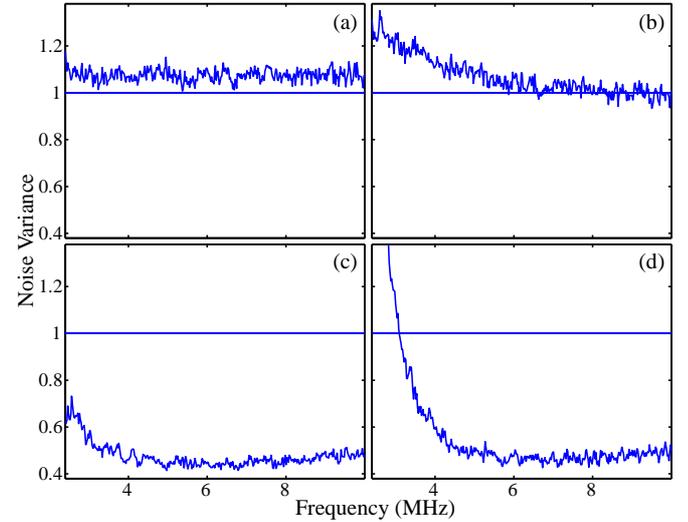

FIG. 8: Experimental measurement of $\Delta^2_{x \pm y}\hat{S}_i$ normalised to the two beam shot noise level. (a) $\Delta^2_{x \pm y}\hat{S}_0$, (b) $\Delta^2_{x \pm y}\hat{S}_1$, (c) $\Delta^2_{x \pm y}\hat{S}_2$, (d) $\Delta^2_{x \pm y}\hat{S}_3$

The variance of the unity gain electronic sum or subtraction of the Stokes operator measurements between the polarisation entangled beams was obtained in a spectrum analyzer at a resolution bandwidth of 300 kHz and video bandwidth of 300 Hz. This resulted in spectra for $\Delta^2_{x \pm y}\hat{S}_i$. These spectra are displayed in fig. 8. From fig. 8(a) and (b) we see that any correlation of $\hat{S}_0$ or $\hat{S}_1$ between the beams is limited by their joint shot noise. On the other hand $\hat{S}_2$ and $\hat{S}_3$ both show correlation to well below the shot noise level.

$\mathcal{I}(\hat{S}_1, \hat{S}_2)$ and $\mathcal{I}(\hat{S}_2, \hat{S}_3)$ were obtained from eqs. (31) and (33), and the measurements of $\Delta^2_{x \pm y}\hat{S}_1$, $\Delta^2_{x \pm y}\hat{S}_2$, $\Delta^2_{x \pm y}\hat{S}_3$, $\alpha_H$ and $\alpha_V$. The resulting spectra are shown in fig. 9. The dashed lines indicate the results a pair of coherent beams would produce. Both traces are below this line throughout almost the entire measurement range, this is an indication that the light is in a non-classical state. At low frequencies both traces were degraded by noise introduced by the relaxation oscillation of our laser. $\mathcal{I}(\hat{S}_2, \hat{S}_3)$ shows polarisation entanglement, however as expected $\mathcal{I}(\hat{S}_2, \hat{S}_3)$ is far above unity. The best entanglement was observed at 6.8 MHz with $\mathcal{I}(\hat{S}_2, \hat{S}_3) = 0.49$.

*Measurement of the EPR criterion*

We determined the EPR paradox criterion in a similar manner to the inseparability criterion. This time, rather than taking the direct unity gain sum or difference between the measurements on beams $x$ and $y$, we took the sum or difference with a gain $g$ chosen to minimise the resulting variances (see eq. 16). Effectively this is asking the question: given a measurement on beam $y$ how well can I infer the value of that variable for beam $x$? Fig. 10 shows the measured conditional variance spectra of beam $x$ for $\hat{S}_2$ and $\hat{S}_3$.

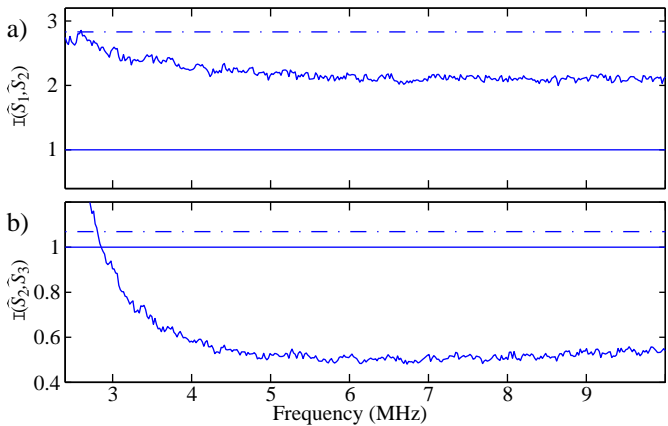

FIG. 9: Experimental measurement of a) $\mathcal{I}(\hat{S}_1, \hat{S}_2)$, b) $\mathcal{I}(\hat{S}_1, \hat{S}_3)$, and c) $\mathcal{I}(\hat{S}_2, \hat{S}_3)$, values below unity indicate entanglement. The dashed line is the corresponding measurement inferred between two coherent beams.

The conditional variance spectra for $\hat{S}_2$ (fig. 10(a)) was obtained from the subtraction of measurements from the two beams, and therefore the relaxation oscillation of our laser, which is strongly correlated between the beams, was almost completely removed. The spectra for $\hat{S}_3$, however, was obtained from the sum of measurements from the two beams, in this case the relaxation oscillation noise is not removed. This resulted in very significant degradation of the conditional spectra for $\hat{S}_3$ (fig. 10(b)).

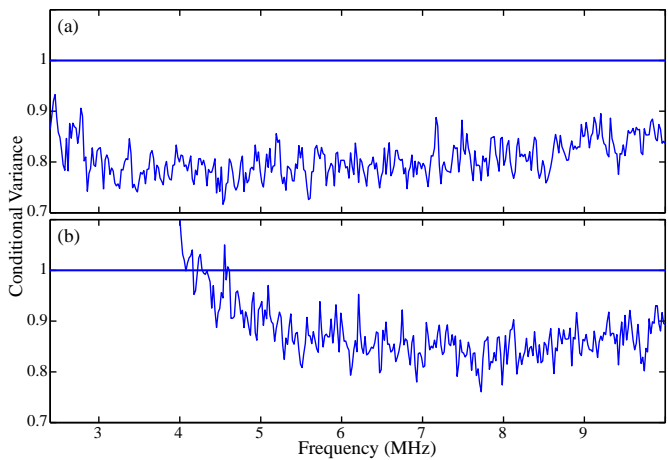

FIG. 10: Experimental measurement of conditional variance a) $\Delta^2_{x|y}\hat{S}_2$, b) $\Delta^2_{x|y}\hat{S}_3$, normalised to the single beam shot noise.

The EPR paradox criterion is the product of the two spectra and is shown in fig. 11. Even with the degradation of the conditional variance of $\hat{S}_3$, the EPR Paradox criterion was verified at frequencies above 4 MHz. The optimum value was $\mathcal{E}(\hat{S}_2, \hat{S}_3) = 0.72$, which was also observed at 6.8 MHz.

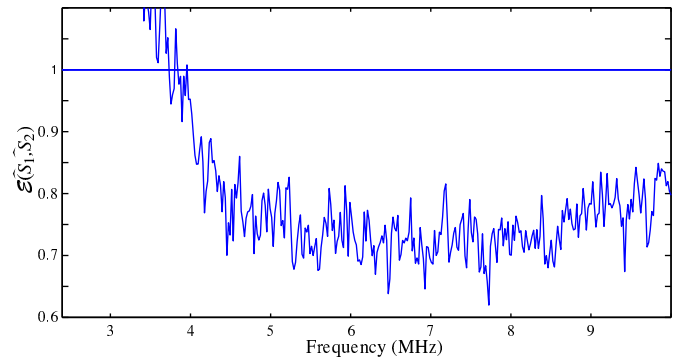

FIG. 11: Experimental measurement of the degree of EPR paradox $\mathcal{E}(\hat{S}_2, \hat{S}_3)$.

It is illustrative to consider our knowledge of beam $x$ before any measurement, and after measurement of $\hat{S}_2$ or $\hat{S}_3$. This can be visualised in terms of noise balls on a Poincaré sphere and is shown in fig. 12. Fig. 12 a) shows the knowledge of the polarisation of beam $x$ at 6.8 MHz without any measurement on beam $y$. $\hat{S}_1$ is shot noise limited, whereas both $\hat{S}_2$ and $\hat{S}_3$ have variances well above the shot noise, as can be seen also in fig. 7. Upon measurement of either $\hat{S}_2$ or $\hat{S}_3$ on beam $y$, that Stokes operator on beam $x$ becomes known to an accuracy below the shot noise (see fig. 12 b) and fig. 12 c) respectively). The product of the uncertainty of $\hat{S}_2$ and $\hat{S}_3$ is then below the uncertainty relation between the two Stokes operators (the dashed circles in fig. 12).

### An explanation of the transformation between quadrature and polarisation entanglement

The Schwinger bosonic representation allows the decomposition of any spin-like operator into a pair of mode operators of the quantum harmonic oscillator [29]. A clearer understanding of the transformation between quadrature and polarisation entanglement presented here may be gained by using this representation to decompose the Stokes operators in terms of quadrature operators. Making assumptions relevant to our experiment discussed earlier ($\alpha_H^2 \ll \alpha_V^2$, $\theta = \pi/2$, and $\langle \delta \hat{X}_H^{\pm} \delta \hat{X}_V^{\pm} \rangle = 0$), we find from eqs. (1), (7) and (8) that the Stokes operator variances of a beam expressed in terms of its horizontally and vertically polarised components are given by

$$\Delta^2 \hat{S}_0 = \alpha_V^2 \, \Delta^2 \hat{X}_H^+ \qquad (37)$$

$$\Delta^2 \hat{S}_1 = \alpha_V^2 \, \Delta^2 \hat{X}_V^+ \qquad (38)$$

$$\Delta^2 \hat{S}_2 = \alpha_V^2 \, \Delta^2 \hat{X}_H^- \qquad (39)$$

$$\Delta^2 \hat{S}_3 = \alpha_V^2 \, \Delta^2 \hat{X}_H^+ \qquad (40)$$

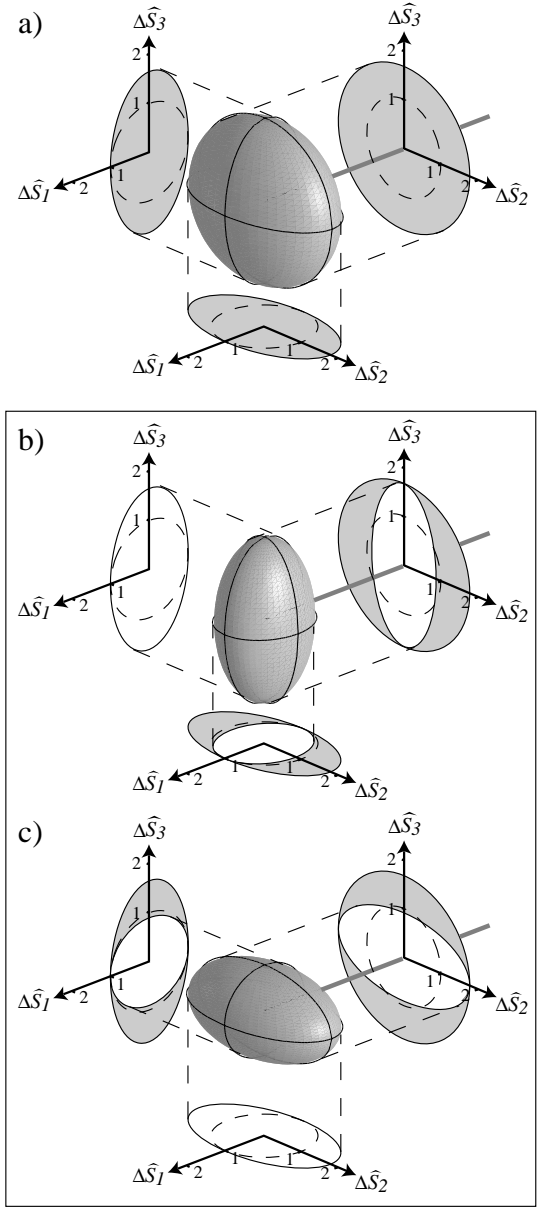

a) Knowledge of the polarisation of beam $x$ before any measurement on beam $y$; b) conditional knowledge of beam $x$ after measurement of $\hat{S}_2$ on beam $y$; c) conditional knowledge of beam $x$ after measurement of $\hat{S}_3$ on beam $y$. The dashed circles define the limit of classical correlation. These representations were generated from results at 6.8 MHz.

The amplitude fluctuations of the bright vertically polarised input beam $X_V^+$ have been mapped on to $\hat{S}_1$ and since $\alpha_H^2 \ll \alpha_V^2$ also onto $\hat{S}_0$. The phase and amplitude fluctuations of the weak horizontally polarised input beam have been mapped onto $\hat{S}_2$ and $\hat{S}_3$, respectively. It is interesting to note that the phase fluctuations of the vertically polarised beam play no role in defining the polarisation of the output. They result in over-all phase fluctuations on the output which have no effect on its polarisation. It is a simple step, then, to derive the sum and difference variances $\Delta_{x \pm y}^2 \hat{S}_i$ between two beams $x$ and $y$

$$\Delta_{x \pm y}^2 \hat{S}_1 = \alpha_V^2 \, \Delta_{x \pm y}^2 \hat{X}_V^+ \tag{41}$$

$$\Delta_{x \pm y}^2 \hat{S}_2 = \alpha_V^2 \, \Delta_{x \pm y}^2 \hat{X}_H^- \tag{42}$$

$$\Delta_{x \pm y}^2 \hat{S}_3 = \alpha_V^2 \, \Delta_{x \pm y}^2 \hat{X}_H^+ \tag{43}$$

It is interesting that any one of these variances can become arbitrarily small without the presence of entanglement. This is achieved simply by squeezing the quadrature of interest on the input modes of both $x$ and $y$. For example, if both horizontal inputs are phase quadrature squeezed then $\Delta_{x \pm y}^2 \hat{S}_2$ can become arbitrarily small. This however, has the natural consequence of $\Delta_{x \pm y}^2 \hat{S}_3$ becoming very large. Interestingly, there is no such consequence for $\Delta_{x \pm y}^2 \hat{S}_1$. Amplitude quadrature squeezing of the vertical inputs allows $\Delta_{x \pm y}^2 \hat{S}_1$ to become arbitrarily small, and has no effect on $\Delta_{x \pm y}^2 \hat{S}_2$ and $\Delta_{x \pm y}^2 \hat{S}_3$. It is therefore possible to make both $\Delta_{x \pm y}^2 \hat{S}_1$ and (say) $\Delta_{x \pm y}^2 \hat{S}_2$ diminishingly small (and therefore their sum also) without any entanglement present. This is the essence of why normalisation of the inseparability and EPR paradox criteria must be performed relative to the uncertainty relation between the operators (eqs. (28-30)) rather than the shot noise of the beams (which is a constant independent of the orientation of the Stokes vector). We see that with $\alpha_H^2 \ll \alpha_V^2$, the uncertainty relation between $\hat{S}_1$ and $\hat{S}_2$ also becomes diminishingly small, so that $\mathcal{I}(\hat{S}_1, \hat{S}_2)$ and $\mathcal{E}(\hat{S}_1, \hat{S}_2)$ (eqs. (31) and (34)) remain greater than unity. In contrast, for quadrature entanglement, the two normalisation procedures are equivalent.

Using eqs. (31-33) and (41-43) the degrees of inseparability $\mathcal{I}(\hat{S}_1, \hat{S}_2)$ and $\mathcal{I}(\hat{S}_2, \hat{S}_3)$, can now be expressed in terms of quadrature sum and difference variances of the input horizontally and vertically polarised modes

$$\mathcal{I}(\hat{S}_1, \hat{S}_2) = \frac{\alpha_V}{\alpha_H} \left( \frac{\Delta_{x \pm y}^2 \hat{X}_V^+ + \Delta_{x \pm y}^2 \hat{X}_H^-}{8} \right) \tag{44}$$

$$\mathcal{I}(\hat{S}_2, \hat{S}_3) = \left( 1 + \frac{\alpha_H^2}{\alpha_V^2} \right) \left( \frac{\Delta_{x \pm y}^2 \hat{X}_H^+ + \Delta_{x \pm y}^2 \hat{X}_H^-}{4} \right) \tag{45}$$

From eq. (44) we see that if the ratio of vertical to horizontal coherent amplitudes $\alpha_V / \alpha_H$ doubles, to retain a given degree of inseparability $\mathcal{I}(\hat{S}_1, \hat{S}_2)$ the level of correlation between $\hat{X}_{V,x}^+$ and $\hat{X}_{V,y}^+$, and between $\hat{X}_{H,x}^-$ and $\hat{X}_{H,y}^-$ must also double. Thus as $\alpha_V / \alpha_H$ increases the level of correlation required for $\mathcal{I}(\hat{S}_1, \hat{S}_2)$ to fall below unity and therefore to demonstrate inseparability quickly becomes experimentally unachievable. In the limit of vacuum horizontal input states ($\alpha_H = 0$) $\mathcal{I}(\hat{S}_1, \hat{S}_2)$ becomes infinite and verification of entanglement is impossible. In contrast, eq. (45) shows that in this situation $\mathcal{I}(\hat{S}_2, \hat{S}_3)$ becomes identical to the criterion for quadrature entanglement (eq. (9)) between the two horizontally polarised inputs. So we see that quadrature entanglement between the horizontally polarised inputs is transformed to polarisation entanglement between $\hat{S}_2$ and $\hat{S}_3$.

The asymmetry between the results for $\mathcal{I}(\hat{S}_1, \hat{S}_2)$ and $\mathcal{I}(\hat{S}_2, \hat{S}_3)$ arises from the the Stokes vector orientation of the two polarizing beam splitter output states. These Stokes vectors are aligned almost exactly along $\hat{S}_1$ (since $\alpha_H^2 \ll \alpha_V^2$). This results in an asymmetry in the commutation relations of eq. (3) and a corresponding bias in the uncertainty relations that define the inseparability criterion.

## POLARISATION ENTANGLEMENT OF ALL THREE STOKES OPERATORS

So far, we have demonstrated polarisation entanglement between two Stokes operators. However, the polarisation of light has three degrees of freedom, and all three can be entangled. This leads to a much more complex quantum state, somewhat analogous to discrete polarisation entanglement between pairs of photons where the correlation is independent of the basis of measurement. In the continuous variable description, however, the mean field polarisation imposes a peculiar basis for the description of polarisation fluctuations which forbids a complete analogy to the discrete case. We will here extend the work of ref. [9] to propose a possible configuration for continuous variable entanglement between all three Stokes operators.

Due to the dependence of the Stokes operator uncertainty relations on their mean fields, symmetric three Stokes operator entanglement requires a symmetric situation for their mean fields. We therefore equate the expectation values of all three Stokes operators for both beams $|\langle \hat{S}_i \rangle| = \alpha^2$. This leads to restrictions on the intensities of the horizontal and vertical states

$$\alpha_V^2 = \frac{\sqrt{3}-1}{2}\alpha^2 \qquad (46)$$

$$\alpha_H^2 = \frac{\sqrt{3}+1}{2}\alpha^2 \qquad (47)$$

and to conditions on the phase relationship between the horizontal and vertical input states

$$\theta_x = \pi/4 + n_x \pi/2 \qquad (48)$$

$$\theta_y = \pi/4 + n_y \pi/2 \qquad (49)$$

where $n_x$ and $n_y$ are integers. The three degrees of inseparability of eqs. (31,32,33) then become symmetric

$$\mathcal{I}(\hat{S}_i, \hat{S}_j) = \frac{\Delta_{x\pm y}^2 \hat{S}_i + \Delta_{x\pm y}^2 \hat{S}_j}{4\alpha^2} \qquad (50)$$

We assume that the two horizontally polarised inputs, and the two vertically polarised inputs, are quadrature entangled with the same degree of correlation such that

$$\Delta_{x\pm y}^2 \hat{X}_H^+ = \Delta_{x\pm y}^2 \hat{X}_H^- = \Delta_{x\pm y}^2 \hat{X}_V^+ = \Delta_{x\pm y}^2 \hat{X}_V^- = \Delta_{x\pm y}^2 \hat{X} \qquad (51)$$

To simultaneously minimise each of $\Delta_{x\pm y}^2 \hat{S}_i$ it is necessary that $\theta_x = -\theta_y + n\pi$. After making this assumption we find that the sum and difference variance between beams $x$ and

$y$ of all three Stokes operators are equal, and can be related directly to $\Delta_{x\pm y}^2 \hat{X}$

$$\Delta_{x\pm y}^2 \hat{S}_1 = \Delta_{x\pm y}^2 \hat{S}_2 = \Delta_{x\pm y}^2 \hat{S}_3 = \sqrt{3}\alpha^2 \Delta_{x\pm y}^2 \hat{X} \qquad (52)$$

Using this relationship, with the polarisation commutation relations of eqs. (3) and the general inseparability criteria eq. (12), the polarisation inseparability criteria can also be directly related to $\Delta_{x\pm y}^2 \hat{X}$

$$\mathcal{I}(\hat{S}_i, \hat{S}_j) = \frac{\sqrt{3}}{2}\Delta_{x\pm y}^2 \hat{X} \qquad (53)$$

and the entanglement is equivalent between any two Stokes operators. The condition for entanglement can then be expressed as a simple criterion on the quadrature entanglement between the input beams

$$\mathcal{I}(\hat{S}_i, \hat{S}_j) < 1 \Longleftrightarrow \mathcal{I}(\hat{X}^+, \hat{X}^-) < \frac{1}{\sqrt{3}} \qquad (54)$$

where $\mathcal{I}(\hat{X}^+, \hat{X}^-) = \mathcal{I}(\hat{X}_H^+, \hat{X}_H^-) = \mathcal{I}(\hat{X}_V^+, \hat{X}_V^-)$. The factor of $\frac{1}{\sqrt{3}}$ arises from the projection of the quadrature properties onto a polarisation basis in which the Stokes vector is pointing at equal angle ($\cos^{-1}(\frac{1}{\sqrt{3}})$) from all three Stokes operator axes. In principle it is possible to have all three Stokes operators perfectly entangled. In other word, ideally the measurement of any Stokes operator of one of the beams could allow the exact prediction of that Stokes operator from the other beam (see fig. 13). However, contrary to the discrete photon case, here the analysis basis orientation is fixed by the mean field polarisation, and rotating this basis would change the commutation relations and therefore the uncertainty relations that verification of the entanglement replies upon. The experimental production of this symmetric polarisation entanglement is a straightforward extension of the experiment reported here, however a significant step-up in resources is required. Four squeezed beams rather than two are required, and due to the projection factor $\frac{1}{\sqrt{3}}$ a higher level of squeezing in each beam is necessary. This entanglement resource would, however, enable the demonstration of maximal continuous variable polarisation teleportation.

## A LOOK AT THE CORRELATION FUNCTION

Throughout this paper we have neglected the correlation function term in the uncertainty relations, this resulted in sufficient, but not necessary, conditions for entanglement. Unlike the commutation relation, the correlation function is state dependent. We would like to look at some of the problems associated with this here, and then examine the values of the correlation functions for the two states discussed specifically in this paper. Again assuming that $\alpha_V \gg 1$ so that higher order terms can be neglected, the correlation functions for each

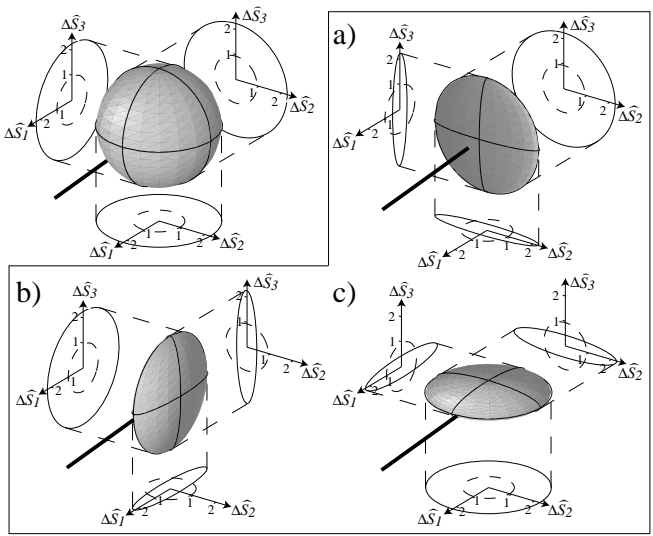

FIG. 13: Calculated polarisation entanglement produced from four pure quadrature squeezed beams with squeezed quadrature variances of 0.1. The top left figure represents the knowledge of beam $y$ before any measurement of beam $x$. a), b), and c) represent the conditional knowledge of beam $y$ given measurements of $\hat{S}_1$, $\hat{S}_2$, and $\hat{S}_3$ respectively on beam $x$. If the conditional knowledge is better than the dashed circles the state is entangled.

of the three permutations of Stokes operators are found to be

$$|\langle \delta\hat{S}_1 \delta\hat{S}_2 + \delta\hat{S}_2 \delta\hat{S}_1 \rangle|^2 = 4\cos^2\theta\,\alpha_H^2\,\alpha_V^2\left(\Delta^2\hat{X}_H^+ - \Delta^2\hat{X}_V^+\right)^2 \quad (55)$$

$$|\langle \delta\hat{S}_1 \delta\hat{S}_3 + \delta\hat{S}_3 \delta\hat{S}_1 \rangle|^2 = 4\sin^2\theta\,\alpha_H^2\,\alpha_V^2\left(\Delta^2\hat{X}_H^+ - \Delta^2\hat{X}_V^+\right)^2 \quad (56)$$

$$|\langle \delta\hat{S}_2 \delta\hat{S}_3 + \delta\hat{S}_3 \delta\hat{S}_2 \rangle|^2 = 4\sin^2\theta\cos^2\theta\left\{\alpha_V^2\left(\Delta^2\hat{X}_H^+ - \Delta^2\hat{X}_H^-\right)\right.$$
$$\left. + \alpha_H^2\left(\Delta^2\hat{X}_V^+ - \Delta^2\hat{X}_V^-\right)\right\}^2 \quad (57)$$

We can, naively, include these correlation functions in the Stokes operator uncertainty relations as shown in eqs. (19); and it seems, obtain more general expressions for the polarisation inseparability criteria $\mathcal{I}_{\text{corr}}(\hat{S}_i, \hat{S}_j)$, of the form

$$\mathcal{I}_{\text{corr}}(\hat{S}_i, \hat{S}_j) = \frac{\Delta_{x \pm y}^2 \hat{S}_i + \Delta_{x \pm y}^2 \hat{S}_j}{2\sqrt{|[\delta\hat{S}_i, \delta\hat{S}_j]|^2 + |\langle \delta\hat{S}_i \delta\hat{S}_j + \delta\hat{S}_j \delta\hat{S}_i \rangle|^2}} \quad (58)$$

Let us consider, for example, $\mathcal{I}_{\text{corr}}(\hat{S}_1, \hat{S}_3)$, and take $\theta = \theta_x = \theta_y = \pi/2$ as in our experiment. Using eqs. (29), (41), (43), and (56) and under the restrictions previous stated (eqs. (20-27) and $\alpha_V \gg 1$) we find that

$$\mathcal{I}_{\text{corr}}(\hat{S}_1, \hat{S}_3) = \frac{\alpha_V}{\alpha_H}\left(\frac{\Delta_{x \pm y}^2 \hat{X}_V^+ + \Delta_{x \pm y}^2 \hat{X}_H^+}{2|\Delta^2\hat{X}_H^+ - \Delta^2\hat{X}_V^+|}\right) \quad (59)$$

The problem with including the correlation term in the inseparability criterion now becomes apparent. Consider that beams $x$ and $y$ are independant coherent states, and therefore clearly separable. Perfectly correlated classical noise can be applied to $\hat{X}_{H,x}^+$ and $X_{H,y}^+$ (or $\hat{X}_{V,x}^+$ and $X_{V,y}^+$) using electronics (classical communication) and amplitude modulators (local operations). It is well know that classical communication and local operations are unable to improve entanglement so, after introduction of this correlated noise, beams $x$ and $y$ must remain separable. By varying the amplitude of the noise (via classical amplification) both $\Delta^2\hat{X}_{H,x}^+$ and $\Delta^2\hat{X}_{H,y}^+$ can be made arbitrarily large. The correlation function between $\hat{S}_1$ and $\hat{S}_3$ (eq. (56)) and hence the general uncertainty product (the denominator of eq. (59)) can then also be made arbitrarily large. Since the noise is perfectly correlated between beams $x$ and $y$ however, it can be arranged to have *no* effect on the sum and difference variances $\Delta_{x \pm y}^2 \hat{S}_1$ and $\Delta_{x \pm y}^2 \hat{S}_3$. Therefore, the numerator of eq. (59) remains unchanged, whilst the denominator may be made arbitrarily large. So as the amplitude of the noise increases $\mathcal{I}_{\text{corr}}(\hat{S}_1, \hat{S}_3) \to 0$, and it appears that beams $x$ and $y$ become entangled. This problem arises because uncertainty product correlation function term implicitly assumes that no information is available about the correlation from an alternative source. Since, the modulation considered here is correlated between the beams, measurements on one beam can provide information about the state of the other, and therefore the uncertainty relations including the correlation function calculated from a single beam are invalid. In order to derive a general necessary and sufficient criteria for polarisation entanglement a much more detailed analysis of the correlation function would be required, we do not present that here.

Let us now consider the two particular states discussed in this paper. In the first case, where entanglement was experimentally observed between $\hat{S}_2$ and $\hat{S}_3$, the relative phase between the horizontal and vertical input modes was controlled to be $\theta = \pi/2$. The correlation functions between $\hat{S}_1$ and $\hat{S}_2$, and between $\hat{S}_2$ and $\hat{S}_3$ (eqs. (55) and (57)) were both equal to zero. Therefore for our experimental configuration the inseparability criteria in eqs. (31) and (33) are necessary and sufficient, and represent a hard boundary between polarisation entangled and non-polarisation entangled states. The entanglement between all three pairs of Stokes operators proposed in section of this paper is highly symmetrical, $\Delta^2\hat{X}_{H,x}^+ = \Delta^2\hat{X}_{H,y}^\pm = \Delta^2\hat{X}_{V,x}^\pm = \Delta^2\hat{X}_{V,y}^\pm$. In this case all three correlation functions can be seen to be zero, and all three inseparability criteria are therefore necessary and sufficient.

## SUMMARY AND CONCLUSION

We have presented the first generation of continuous variable polarisation entanglement. This entanglement is achieved by transformation of the well know and well understood quadrature entanglement onto a polarisation basis. In order to characterise the entanglement we generalise both the inseparability criterion of Duan *et al* [1], and the EPR paradox criteria of Reid and Drummond [2]. We utilise the standard uncertainty relation in this generalisation, which results in sufficient polarisation entanglement criteria. We briefly consider the generalised uncertainty relation, which includes the correlation function. We demonstrate that the correlation function is zero for the two situations considered in this paper, and therefore the inseparability and EPR paradox criteria derived are, for these cases, necessary and sufficient. A form of the criteria that is, in general, necessary and sufficient, is a problem for future analysis. In our experiment both criteria were observed to be well inside the regime for entanglement between the Stokes operators $\hat{S}_2$ and $\hat{S}_3$.

We have shown that with an available resource of four quadrature squeezed beams it is possible for all three Stokes operators to be perfectly entangled, although with a bound $\sqrt{3}$ times lower (stronger) than that for quadrature entanglement.

This work is supported by the Australian Research Council and is part of the EU QIPC Project, No. IST-1999-13071 (QUICOV). R. S. acknowledges the Alexander von Humboldt foundation for support.